\documentclass[%
superscriptaddress,
preprint,
bibnotes,
amsmath,amssymb,amsfonts,amsthm,amscd,bbm,
aps,
pra,
floatfix,
]{revtex4-1}

\usepackage{graphicx}
\usepackage{dcolumn}
\usepackage{bm}
\usepackage{listings}
\usepackage{color}

\newif\ifNOSUP \NOSUPfalse

\begin{document}


\title{Touching proteins with virtual bare hands: how to visualize protein-drug complexes and their dynamics in virtual reality}

\author{Erick Martins Ratamero}
\affiliation{
Department of Physics, University of Warwick, Coventry, CV4 7AL, UK
}

\author{Dom Bellini}
\affiliation{
School of Life Sciences, University of Warwick, Coventry, CV4 7AL, UK
}

\author{Christopher G.\ Dowson}
\affiliation{
School of Life Sciences, University of Warwick, Coventry, CV4 7AL, UK
}

\author{Rudolf A.\ R\"{o}mer}
\email{R.Roemer@warwick.ac.uk}
\affiliation{
Department of Physics, University of Warwick, Coventry, CV4 7AL, UK
}

\date{$Revision: 1.0 $, compiled \today}

\begin{abstract}
The ability to precisely visualize the atomic geometry of the interactions between a drug and its protein target in structural models is critical in predicting the correct modifications in previously identified inhibitors to create more effective next generation drugs. It is currently common practice among medicinal chemists while attempting the above to access the information contained in three-dimensional structures by using two-dimensional projections, which can preclude disclosure of useful features. A more precise visualization of the three-dimensional configuration of the atomic geometry in the models can be achieved through the implementation of immersive virtual reality (VR). In this work, we present a freely available software pipeline for visualising protein structures through VR. New customer hardware, such as the {\sc HTC Vive} and the {\sc Oculus Rift} utilized in this study, are available at reasonable prices. Moreover, we have combined VR visualization with fast algorithms for simulating intramolecular motions of protein flexibility, in an effort to further improve structure-lead drug design by exposing molecular interactions that might be hidden in the less informative static models.   
 
\end{abstract}



\maketitle
%
%

\section{\label{sec:level1}Introduction}
Proteins are three-dimensional (3D) objects  
%
\cite{Riddell1974} and, for the last century, spatially-resolved structural models of proteins and other biologically relevant molecules have been provided by various experimental techniques. 
X-ray crystallography was particularly instrumental in this revolution \cite{Anderson2003} with the very first structure of a protein resolved by this method in the 1950s \cite{Kendrew1958}. Since then, X-ray crystallography has led to the building of detailed protein models and was instrumental for a number of important advances, with Watson and Crick's accurate 3D model of the DNA structure as a prominent example \cite{Watson1969}.
The 3D characteristics of protein molecules are important in aiding our comprehension of many biological processes. Furthermore, beyond the classical "structure implies function" approach, it is now also becoming increasingly clear that protein dynamics is key to understanding protein function \cite{HenK07}. One of the ways we could potentially access this information is by interacting with, manipulating and visualising static and dynamic models of such proteins in 3D. These might be constructed as real objects or exist in a virtual reality (VR) environment.
A large part of scientific and medicinal research on drugs, such as, e.g., understanding antimicrobial resistance (AMR), revolves around clarifying the ways drugs bind to proteins and vice versa. In the, e.g., AMR context, viewing protein structures and their dynamics and understanding how mutations can lead to conformational changes and, thus, changes in binding regions that are relevant to AMR, is essential. 

Since computers became ubiquitous, the development of \emph{computer models} for proteins, as opposed to physical ones, has became progressively easier and cheaper. With that came the possibility to show proteins in stereoscopic 3D view. Many such projects were developed, as TAMS, for example, which used polarized slide projectors to display stereoscopic 3D images \cite{Feldmann1980}.
Many tools exist to make viewing protein structures possible, such as {\sc PyMol} \cite{Pymol}, {\sc VMD} \cite{Humphrey1996}, {\sc Rasmol} \cite{Sayle1995} and {\sc Chimera} \cite{Pettersen2004}. In recent years, there was a further push to develop systems based on web browsers, such as {\sc iView} \cite{Li2014} and {\sc Jmol} \cite{Jmol:3D}. To some degree, every one of those tools can produce stereoscopic 3D images, be it through passive 3D (using chromatic distortion glasses), active 3D (with shutter glasses synchronised to the image displaying device) and autostereoscopic 3D (no headgear required) \cite{McIntire2014}. Though their method changes significantly, they all aim for the same near-3D effect for the end user.
Besides technical drawbacks that either limit resolution or require expensive equipment, these attempts at providing 3D perception when analysing protein structures lack \emph{immersion} into a different environment and true 3D depth perception. 

VR allows us to address this lack of immersion, and introduce a level of interaction between user and visualisation tool that was not possible before. While the usage of VR in research is not new, the current levels of performance with affordable cost definitely are.
Implementing a VR can be achieved through many different techniques. On one side of the complexity scale, we could point towards whole-room arrangements like CAVEs (cave automatic virtual environments) \cite{Creagh}, while extremely simple solutions like Google Cardboard \cite{GoogleVR} would lie at the other end of that scale. Somewhere in between we find modern head-mounted devices (HMDs), like the {\sc Oculus Rift} \cite{OculusRift} and the {\sc HTC Vive} \cite{VIVE}. These create a stereoscopic 3D effect through the usage of LCD displays and lenses that allow the system to display different images for each eye, with slightly different points of view that mimic the position of the eyes of a virtual observer. These HMDs are relatively easy to use and to program, with excellent display quality and affordable prices. 
The prominence of such HMDs amongst the gaming community is particularly useful for researchers. Since VR gaming has gained popularity, the tools for programming software to use HMD capabilities have become better, more streamlined and easier to use for people even without a background in graphics programming. Initiatives such as {\sc SteamVR} \cite{SteamVR} and VR addons for the {\sc Unity3D} game engine \cite{UnityReality} decrease the amount of work necessary to build a VR application considerably.

This paper reports on a protocol for introducing protein structures into VR programs, using a combination of widely and freely available software and custom-built scripts and programs. Besides being an useful tool for researchers to visualise conformational changes in proteins, VR also provides a great outreach opportunity to motivate the general public to understand proteins and their relationship to biochemical research and its applications in drug discovery better. We aim with this paper to give our readers the necessary tools to start their own VR applications. Freely available VR setups for {\sc HTC Vive} and {\sc Oculus Rift} can be downloaded from Ref.\ \cite{MartinsRatamero2017}.

\section{\label{background}Background and Motivation}

Exploring VR environments is not a recent pursuit. The Sensorama \cite{HeiligMorton1961}, first proposed in 1955 and constructed in 1962, is one of the earliest examples of such attempts. The late 1980s was the first period when VR applications and technology drew real interest, with fast growth in adoption during that time. However, the display technology lacked definition, and most products were dismissed as being low quality and unresponsive, both in terms of response time and tactile feedback \cite{Briggs1996TheReality}. 
The responses to these problems were products with exorbitant costs that were not accessible for many research groups and most consumers. As inaccessible as they were, these products created excitement around VR, while the technology was still not ready to deliver results for most people, and the interest in VR decreased towards the end of the 1990s. 
At the same time, the usage of video game technology for research and clinical purposes ("serious games") started gaining popularity \cite{Stone2009}. This movement drove a resurgence in interest for VR technologies; as an example, visualisation of volumetric data \cite{Laha2012} and engineering assemblies \cite{Seth2011} were two of the subjects of VR development in recent years. 
The current VR "boom" differs from the one during the 80's and 90's. This time, consumer-level hardware is relatively affordable and presents high-quality displays, and game software technology is sufficiently developed to make creating new VR applications simple and fast, even for scientists without training in graphics programming.

In our previous work on AMR, we have developed methods and algorithms for simulating the mobility of proteins based on flexibility \cite{Romer2016c}. This method requires four orders of magnitude less computing power when compared with molecular dynamics, and yields a series of conformers that describe the large-scale motions of a protein.
During this project, it became immediately apparent how difficult it was to distinguish between consecutive conformers and visualise the difference between them. It was, then, not easy to see which motions were actually taking place in the protein structure, and near impossible to derive any biological insight from that.
Our initial approach was to generate videos from the series of conformers. This approach clarified the kinds of motion that were happening in larger time scales, and it allowed us to make use of our flexibility-based simulations to better understand the interactions of the proteins being studied.
However, it was still difficult to identify motions in specific residues and domains of the proteins. Specific points of view were necessary for generating videos, and those would always make certain areas clearer, while others were obscured. In this context, an interactive solution where the protein appears as an animated 3D object that can be freely rotated and manipulated is desirable.
Thus, VR emerges as an ideal solution for visualising protein dynamics. It allows us to clearly communicate the large-scale motions taking place on the protein structure, while making interaction possible for looking closer at specific areas of interest, such as binding sites. 

\section{\label{workflow}Workflow}

Our choice for an HMD for VR is the {\sc HTC Vive} \cite{VIVE}. It uses two infrared emitters (``lighthouses") in the corners of the play area, which generate infrared beams in sweeping patterns. The headset possess multiple infrared photo diodes, which will detect these beams, and the position of the headset in the room can be reconstructed from the time differences between signals received at each diode \cite{Dempsey2016}. The user can move around the room freely, with the only inconvenience being the cables attached to the headset. 
Furthermore, the {\sc HTC Vive} can also track controllers to grab and manipulate objects in virtual environments, and can be precisely tracked and displayed inside the virtual environment as well. This allows for an immersive, interactive experience, where the limits are only the HMD cable extension and the corners of the delimited play area defined by the position of the lighthouses.
\lstdefinestyle{customc}{
  belowcaptionskip=0.5\baselineskip,
  breaklines=true,
  frame=single,
  xleftmargin=\parindent,
  language=C,
  numbers=left,
  showstringspaces=false,
  basicstyle=\scriptsize\ttfamily,
  keywordstyle=\bfseries\color{green},
  commentstyle=\itshape\color{red},
  identifierstyle=\color{blue},
  stringstyle=\color{black},
}
\lstset{escapechar=@,style=customc}

\begin{figure}[tbh]
\begin{minipage}{0.85\textwidth}
\begin{lstlisting}
using System.Collections;
using System.Collections.Generic;
using UnityEngine;

public class prot_animator : MonoBehaviour {
    public GameObject fatherObject;
    List<Transform> childrenObjects;
    int childrenCount;
    public float framerate = 30.0f;
    int frameCounter = 0;
    bool goingForwards = true;
	void Start () {
        childrenCount = CountChildren();
        childrenObjects = GetChildren();
        InvokeRepeating("UpdateAnimation", 2.0f, 1.0f / framerate);
	}

    void UpdateAnimation()
    {
        if (frameCounter == childrenCount-1)
        { goingForwards = false;}
        if (frameCounter == 0)
        { goingForwards = true;}
        
        if (goingForwards)
        {
            childrenObjects[frameCounter].gameObject.SetActive(false);
            childrenObjects[frameCounter + 1].gameObject.SetActive(true);
            frameCounter = frameCounter + 1;
        }
        else
        {
            childrenObjects[frameCounter].gameObject.SetActive(false);
            childrenObjects[frameCounter - 1].gameObject.SetActive(true);
            frameCounter = frameCounter - 1;
        }
    }
\end{lstlisting}
\end{minipage}
\caption{\label{unitycode}C\# code for {\sc Unity3D}, a part of the script written to animate proteins. There is a father object with multiple children objects containing the 3D models of consecutive frames of animation, and this script replaces the object being displayed multiple times per second.}
\end{figure}
Our application uses {\sc Unity3D}, a programming and execution environment \cite{UnityReality}. This is a game engine, which allows us to implement ideas quickly and easily. Since it provides standard VR features such as graphics, physics (gravity and rigid-body modelling) and lighting, people without programming experience in these areas can also develop their ideas with minimal training involved. Whenever the standard features are not enough, {\sc Unity3D} also allows us to write scripts (in C\#, in our case) to treat special cases or implement features that are not readily available. One of these examples is presented as Fig. \ref{unitycode}. Here, we have written some code to implement protein animation by replacing the structure being presented multiple times per second.
For interfacing with the virtual reality equipment, we use the {\sc SteamVR} abstraction layer. By using such a tool, it is possible to create a single project that works seamlessly on the two most popular HMDs ({\sc Oculus Rift} and {\sc HTC Vive}), without any code duplication.
However, using such a tool for creating our applications brings downsides as well. Not every file format can be easily imported into {\sc Unity3D}; specifically, importing 3D models with correct colours can be tricky. Since protein structures will be represented as 3D models, this required a specific set up to allow us to visualise these structures inside {\sc Unity3D}.

\begin{figure}[tb]
\includegraphics[height=0.3\textheight]{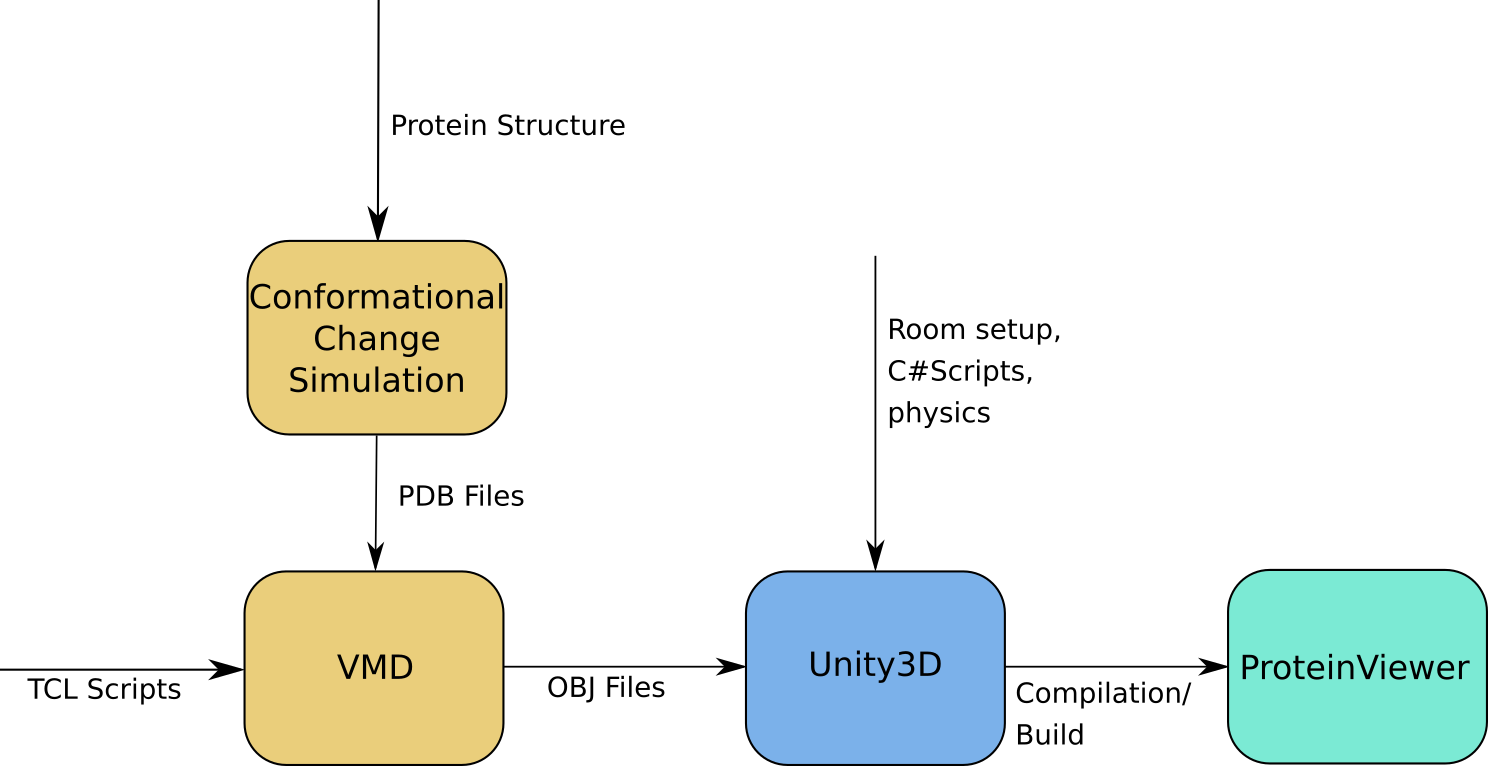}
\caption{\label{diagram}Workflow diagram. Conformational changes for a protein PDB file are simulated, the snapshots are turned into 3D models by {\sc VMD} and then imported into {\sc Unity3D}, which builds the final application called {\sc ProteinViewer}.}
\label{fig:unityscript}
\end{figure}

We present our current workflow for displaying protein structures in VR in Fig.\ \ref{diagram}. The initial inputs to our application are Protein Data Bank (PDB) \cite{Bernstein1977} files. These are text files that describe the geometric structure of molecules. It allows for description of atomic coordinates, rotamers, secondary structures and connections between atoms. In our case, we are mainly concerned with the 3D position of each atom in the protein at each simulation time step; these positions need to be translated to an animated 3D protein structure.
Representing individual atoms in a protein does not indicate clearly 3D features such as twist and fold. Therefore, we will use the ribbon diagram representation to visualise secondary structures. These are representations where the polypeptide backbone is interpolated by a smooth curve, generating helices, sheets and loops. The result is a simple, yet informative diagram where 3D information about the protein can be easily displayed. In our case, we use the Visual Molecular Dynamics (VMD) program \cite{Humphrey1996} for creating representations. Other representations such as the full "sphere" atom representation can of course be chosen as well if desired.
The next step is generating 3D models for the ribbon representations. VMD allows us to create OBJ files; this is an open file format for 3D geometry, storing information about individual vertices, vertex normals, faces of polygons and textures associated with them. This way, we can export the 3D ribbon representations (and any other 3D structure, for that matter) from VMD while keeping any visual information such as colouring associated with it.
Finally, it is necessary to import the generated OBJ files into {\sc Unity3D}. Fortunately, {\sc Unity3D} imports OBJ files natively, and integrates textures without any extra effort being necessary. Therefore, the VMD-generated structures can immediately become physical objects in the VR environment. {\sc Unity3D} allows us to attach collider objects and rigid-body physics to the protein objects, integrating them into the physics framework of the application seamlessly.

It is also possible to introduce dynamic structures into VR. We have use data generated by our simulation method \cite{Jimenez-Roldan2012} in the form of multiple PDB files acting as snapshots of the protein conformation over time. Next, we load each individual PDB file into our pipeline, obtaining 3D models for each frame. We then use a script in {\sc Unity3D} to flip through the models over time (cp.\ Fig.\ \ref{fig:unityscript}), creating an animation that shows the conformational changes calculated by our simulations.

\section{\label{casestudies} Case Studies}

\begin{figure}[tb]
(a)\includegraphics[width=0.45\textwidth]{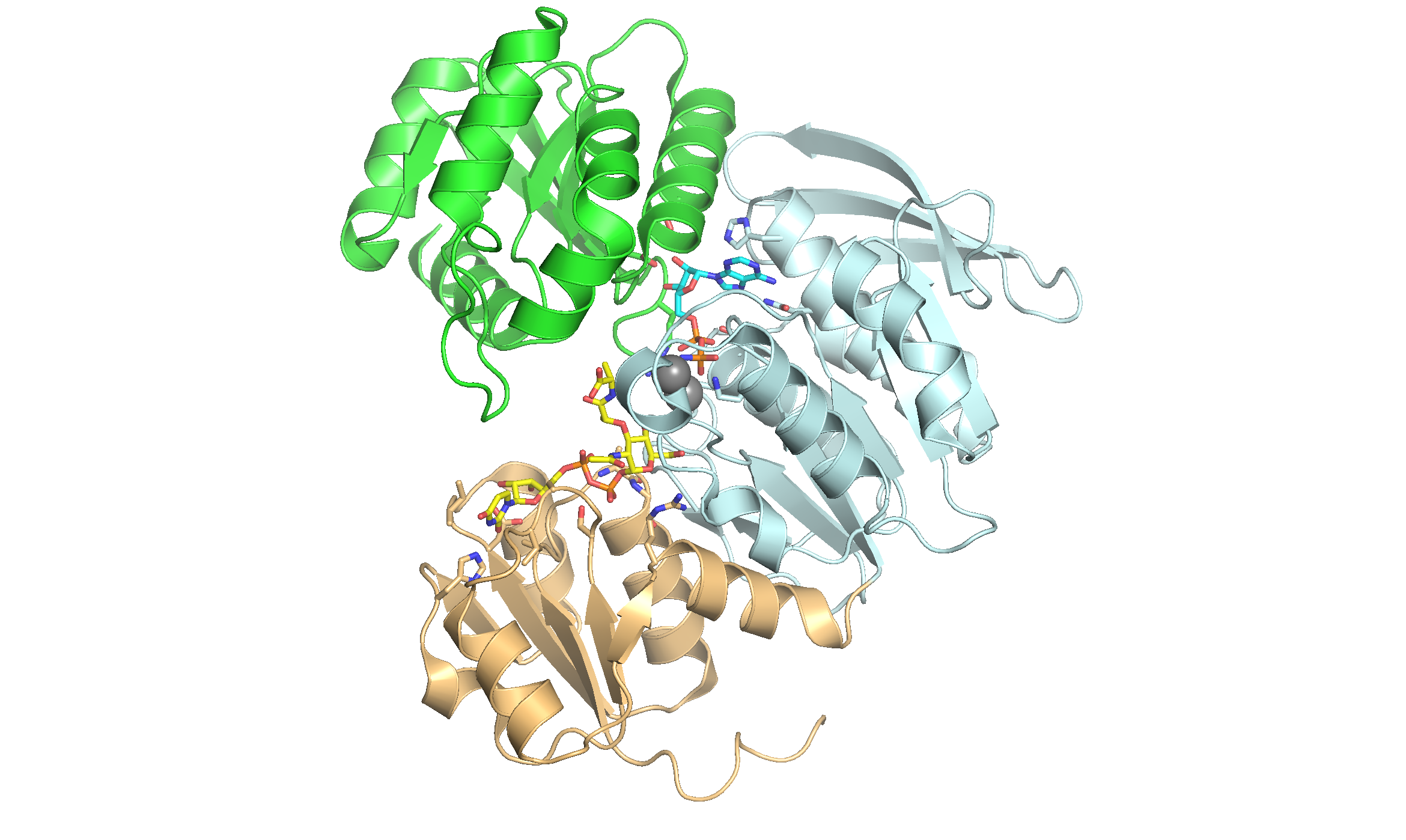}
(b)\includegraphics[width=0.45\textwidth]{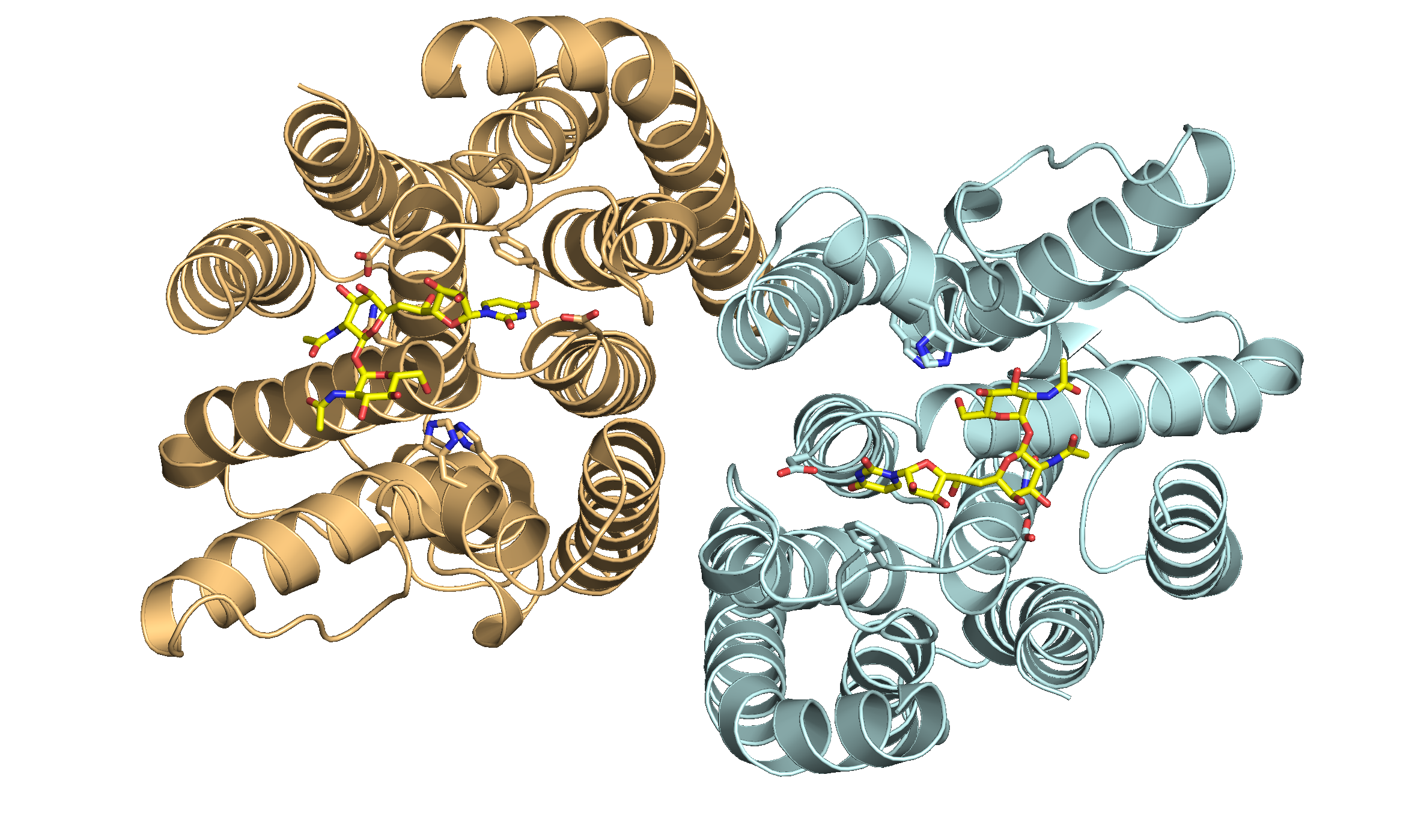}\\
(c)\includegraphics[width=0.45\textwidth]{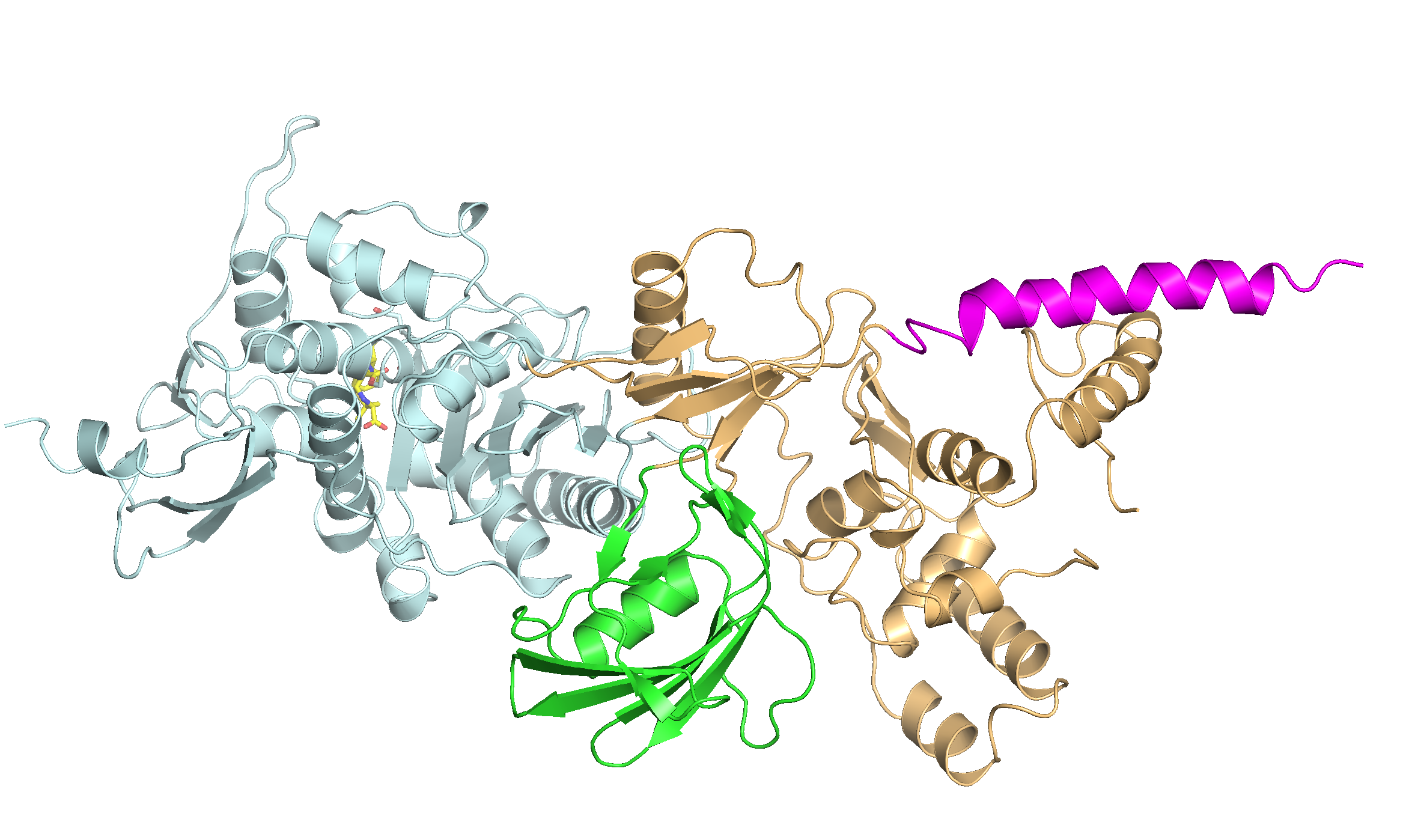}
(d)\includegraphics[width=0.45\textwidth]{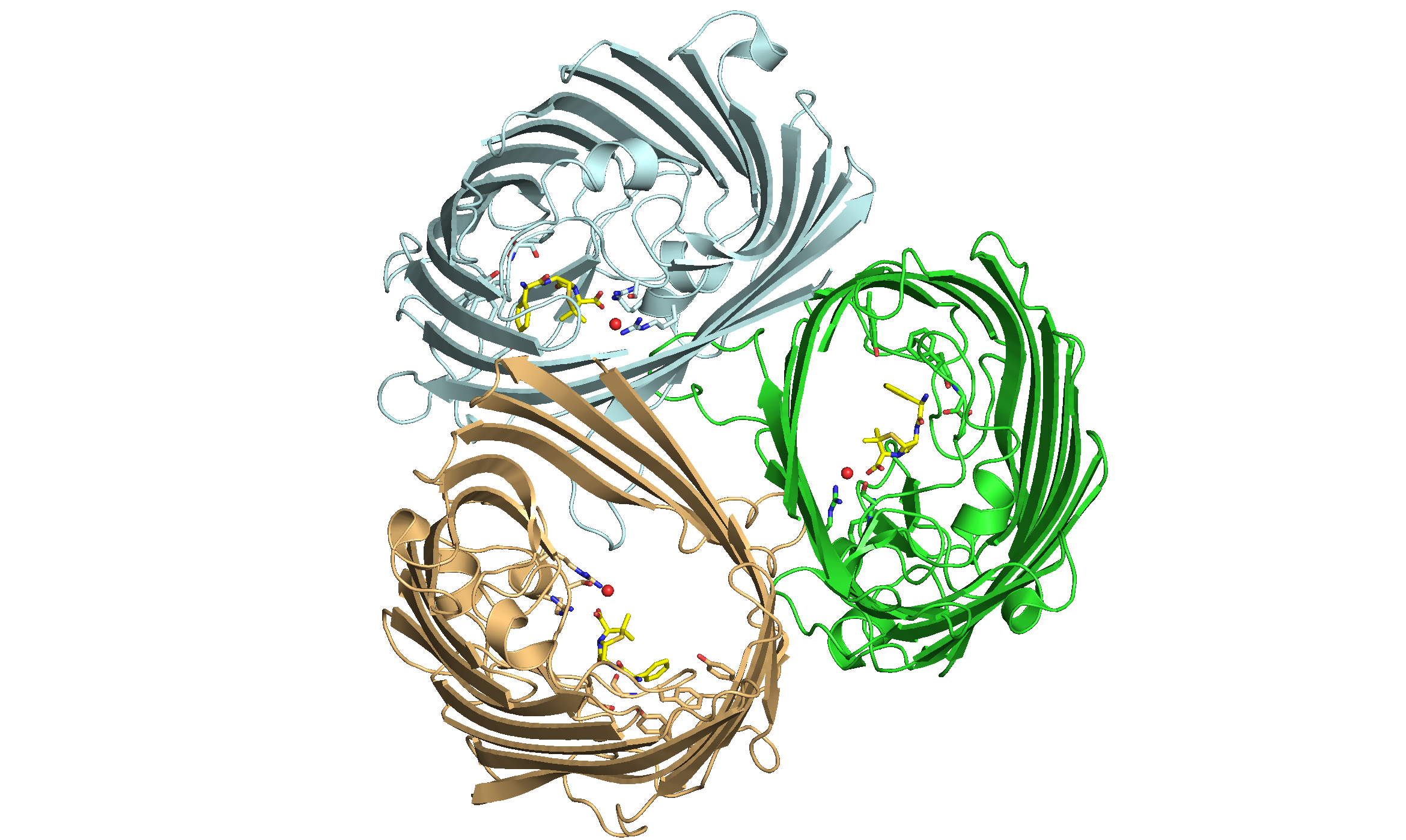}
\caption{
Standard secondary structure visualization using the ``cartoon" option in {\sc Pymol}. Different colors highlight either different domains (a and c) or chains (b and d) of the proteins, while the "stick" representation has been used to indicate ligands and nearby interacting side-chains for:
(a) monomeric MurC with bound substrate UDP-N-acetylmuramoyl-L-alanine, non-hydrolyzable gamma-Imino-ATP and two manganese atoms (PDB code 1P3D),
(b) dimeric MraY in complex with tunicamycin (PDB code 5JNQ),
(c) monomeric PBP1B in complex with TG domain inhibitor, moenomycin, and TP inhibitor, ampicillin (unpublished data) and
(d) trimeric OmpF in complex with ampicillin (PDB code 4GCP).
}\label{mray_struct}
\label{murc_struct}
\label{ompf_struct}
\label{pbp1b_struct}
\end{figure}

The structural models used in this study for the implementation of 3D visualization in VR are those of four bacterial proteins from different cellular compartments: cytoplasm, inner membrane, periplasm and outer membrane. 
MurC is a cytoplasmic ligase involved in the construction of the pentapeptide stem, which is a central component of the peptidoglycan (Fig.\ \ref{murc_struct}a). MraY is an integral cytoplasmic membrane enzyme that catalyzes the formation of lipid II by transferring the pentapeptide to the lipid carrier, which is another essential step in peptidoglycan biosynthesis (Fig.\ \ref{mray_struct}b). Penicillin-binding protein 1b (PBP1b) is a bifunctional enzyme containing both a transglycosilation (TG) and a transpeptidation (TP) domain, which is able to polymerize lipid II to form the peptidoglycan mesh (Fig.\ \ref{pbp1b_struct}c). OmpF is an integral outer membrane channel exploited by most antimicrobial drugs to enter the organism on their way to the target (Fig.\ \ref{ompf_struct}d).
All structures are in complex with either the substrate or antimicrobial drugs that inhibit bacterial cell wall (peptidoglycan) synthesis. As such, they are currently studied targets for understanding AMR resistance mechanisms \cite{BugBDR11}.



Initially, intrinsic motions of these proteins were simulated as presented above by analysing protein flexibility, defining movement modes through elastic network modeling and generating conformers based on that information and steric interactions. These simulations produced a series of PDB files detailing the state of protein structures at certain "snapshot" moments (e.g., every 100th conformer out of $5000$ overall for each mode of motion). 
The output of these simulations are PDB files containing the motions of protein atomic coordinations over time. Unfortunately, the PDB format is not directly supported by the VR software {\sc Unity3D}. Therefore, {\sc VMD} was subsequently used to transform PDB files into OBJ files. TCL scripts \cite{TCL17} for VMD were written to go through output folders from the simulations and automatically generate OBJ files of each conformer. When writing these TCL scripts, care was taken to ensure that positive and negative directions of motion blend together seamlessly. An example of such script for a specific protein is presented as Fig.\ \ref{tclcode}. Here, every PDB file from folders with the ``pos'' and ``neg'' tags for positive and negative directions of protein movement gets loaded onto VMD, specific colouring and representation options for that protein are chosen and finally each frame is rendered and stored as a Wavefront (OBJ) file. A dataset containing all source code used in this project is available in a public repository for download \cite{MartinsRatamero2017}.
Through this process, a series of OBJ files are obtained that act as 3D ``frames" of animation. Upon importing these files into {\sc Unity3D}, the only task left is to ensure that these frames are shown sequentially to impart upon the user the illusion of an animated protein. In this study we do so by updating the displayed object every $1/30$ of a second. Though theoretically possible, it was decided against updating the collision structure of the object at each frame, as this would consume significant computational resources, even if it would allow for ``correct" physics in VR at all times. Instead, we have chosen to define a bounding box spanning the whole range of movement for the protein as a constant collision structure.
\begin{figure}[tbh]
\lstset{language=tcl}
\begin{minipage}{0.85\textwidth}
\begin{lstlisting}
set frame  0
axes location off
set folder [lindex $argv 0]
set negfolder $folder*-neg/*.pdb
set posfolder $folder*-pos/*.pdb
set outputfolder $folder/outputs/animate.%04d.obj

foreach pdb [lsort -decreasing [glob $posfolder]] { 
  mol addfile $pdb 
  incr frame
} 

foreach pdb [lsort [glob $negfolder]] { 
  mol addfile $pdb 
  incr frame
} 

mol modselect 0 0 all
mol modstyle 0 0 Newcartoon
mol modcolor 0 0 Chain
color Chain A orange3
color Chain B mauve
color Chain C lime
mol addrep 0
mol modselect 1 0 not protein
mol modstyle 1 0 licorice
mol modcolor 1 0 Element
color Element C cyan3

for {set framecount 0} {$framecount < $frame} {incr framecount 1} {
	set filename [format $outputfolder $framecount]
	animate goto $framecount
	render Wavefront $filename
}
\end{lstlisting}
\end{minipage}
\caption{\label{tclcode}{\sc TCL} code for {\sc VMD}. Here, we present the script written to generate multiple OBJ files in the correct order to be imported into {\sc Unity3D}. We load all PDB files from specific folders, colour them accordingly for the specific protein structure at play and render them as OBJ files.}
\end{figure}

A ``template" room was constructed for visualising proteins structures in {\sc Unity3D}. This saves time when a quick visualisation is necessary by importing the relevant OBJ files into an existing ``template" project where lighting and physics of the virtual room have previously been defined, and into which VR structures have already been hooked. Use of a ``template" project reduces the tasks to simply importing OBJ files to create ``father" objects for the 3D models and to define rigid body physics and collider boxes. If necessary, a script can be attached to that object for animation. Depending on the number of snapshots in PDB files, the process from simulation outputs to ready-to-use VR application can be as short as 5 minutes.

\begin{figure}[tb]
(a)\includegraphics[width=0.45\textwidth]{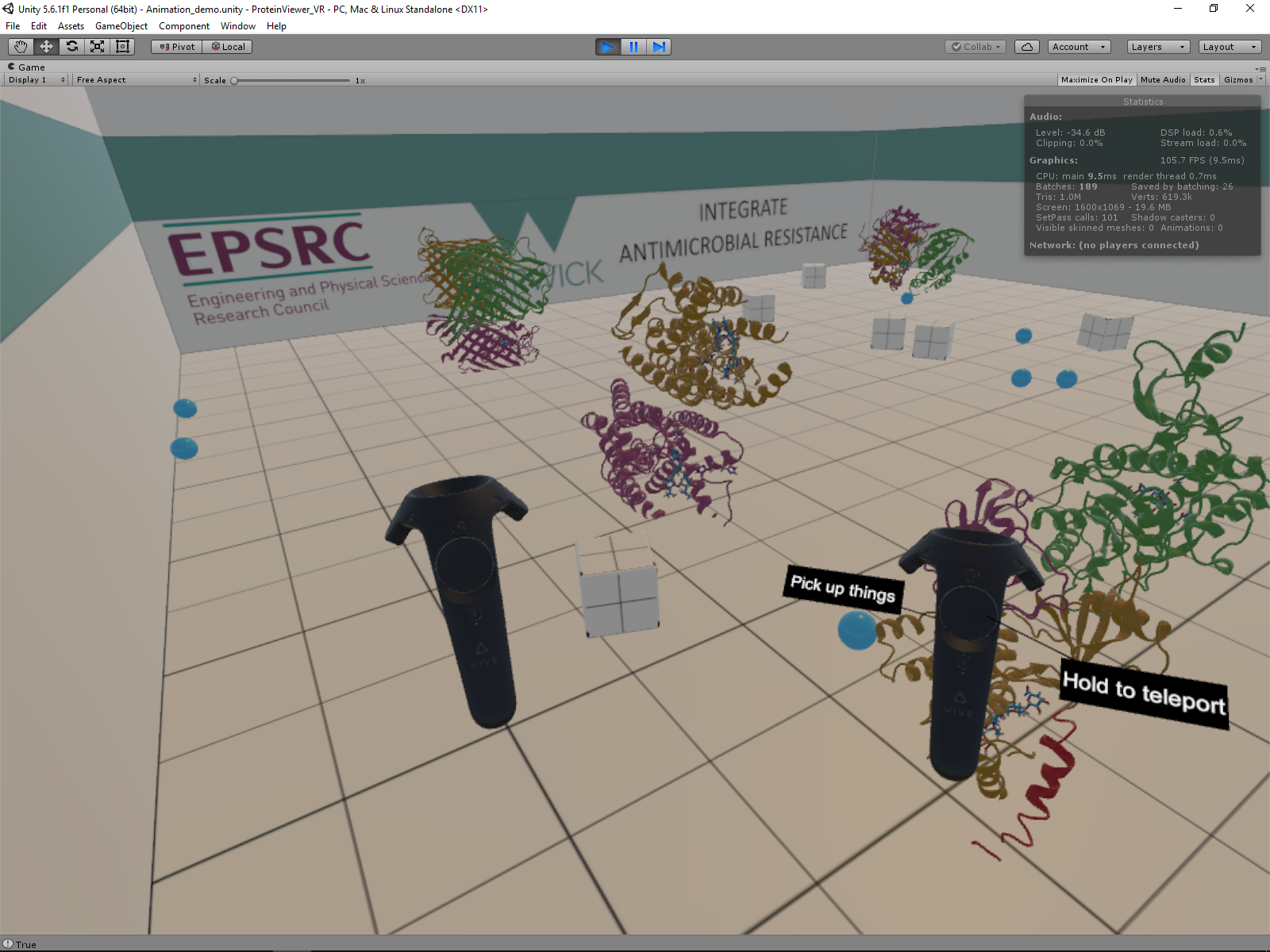}
(b)\includegraphics[width=0.45\textwidth]{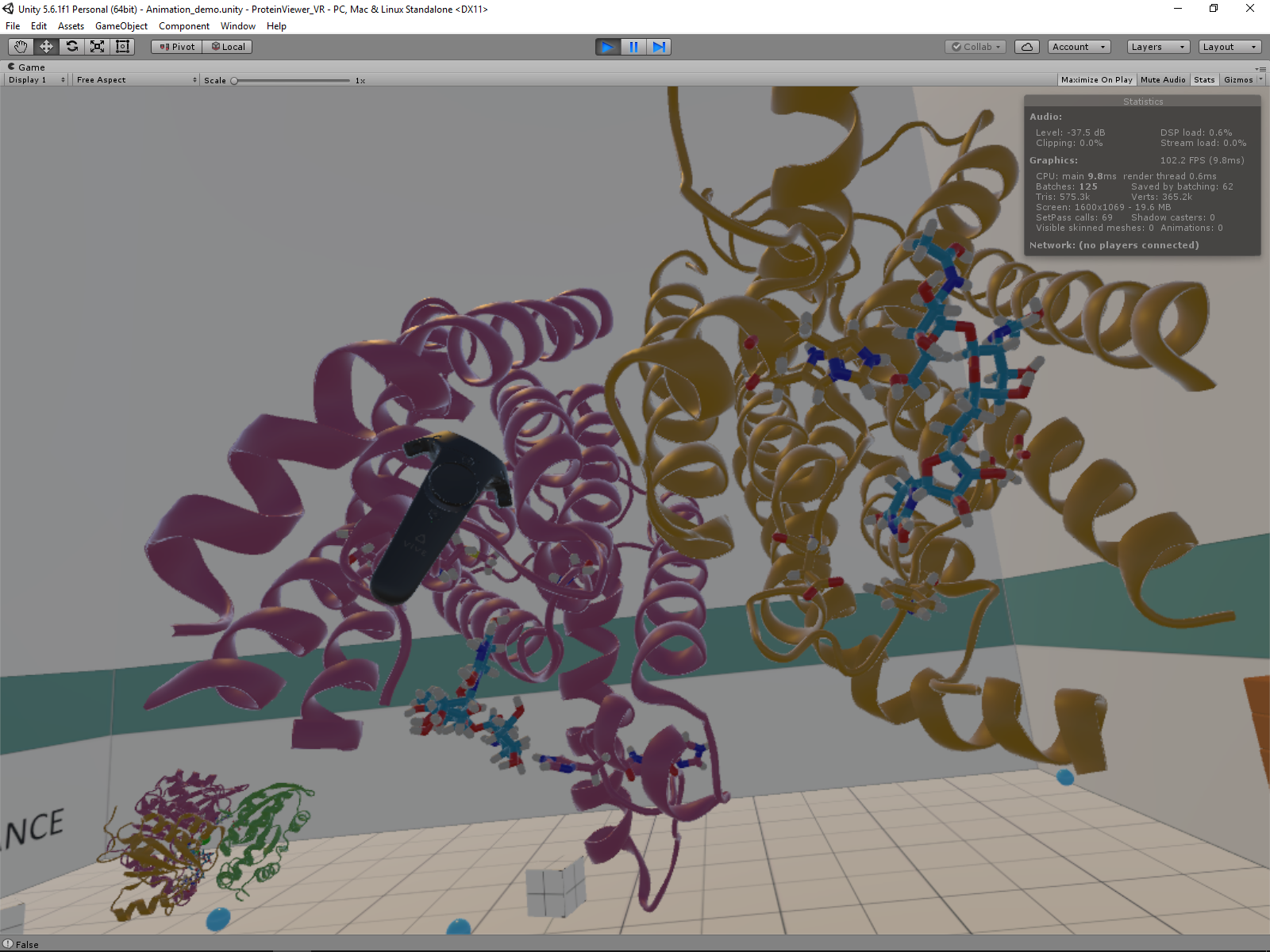}
(c)\includegraphics[width=0.45\textwidth]{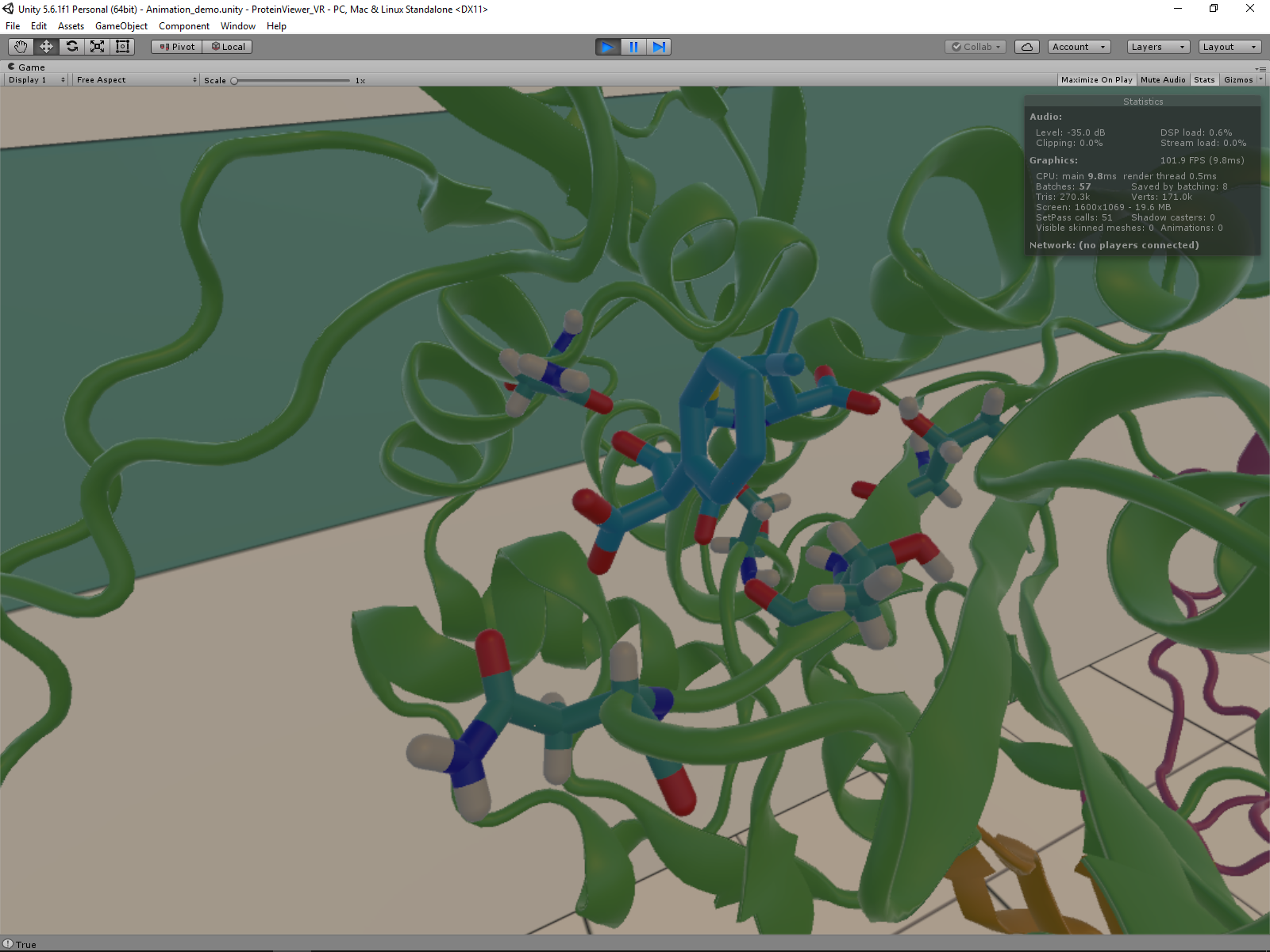}
(d)\includegraphics[width=0.45\textwidth]{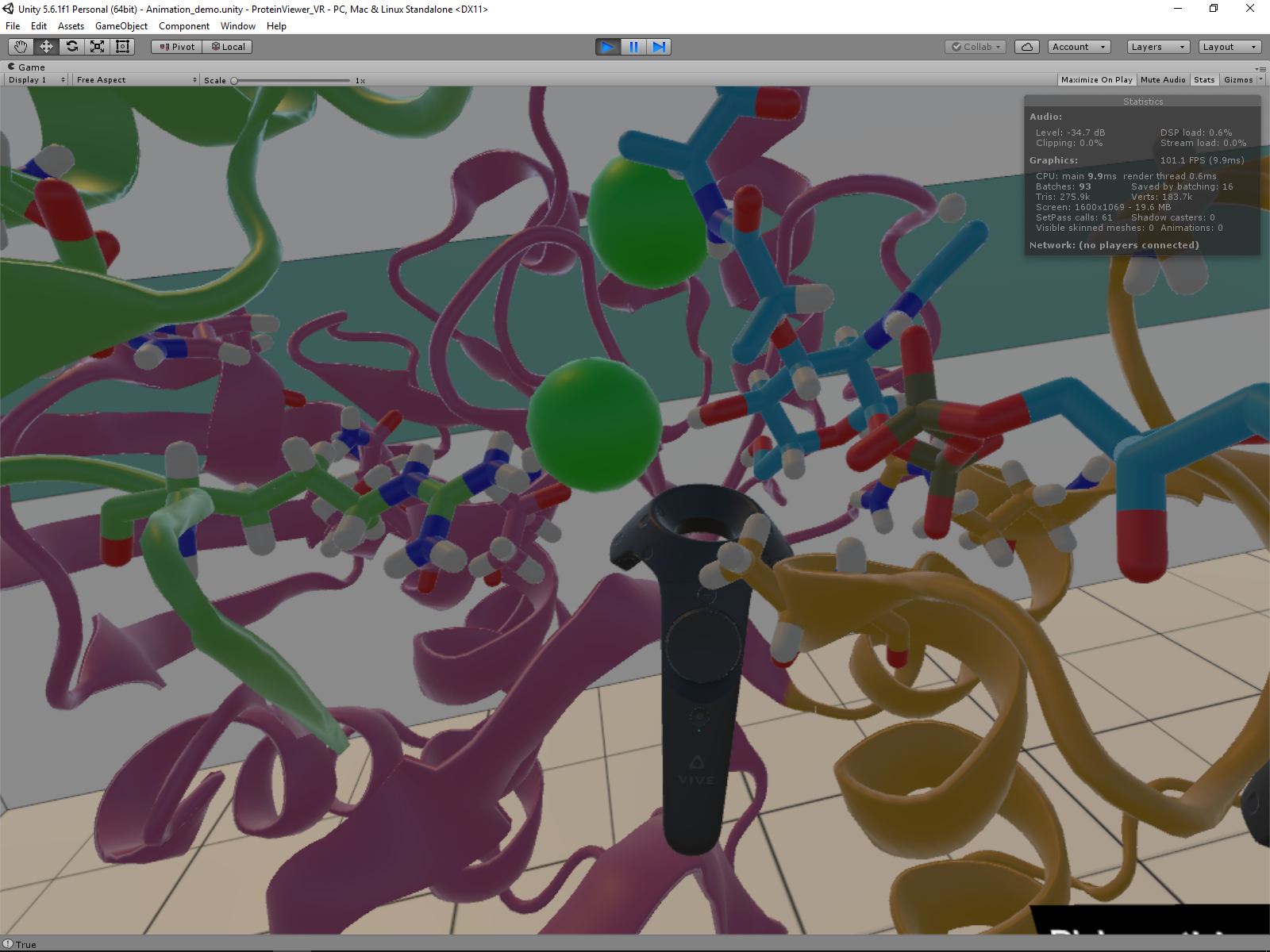}
\caption{%
(a) Template VR room containing the four protein structures discussed above. Both controllers are visible, with tool tips for teaching the controls.
(b) MraY structure in VR. Details of specific residues and drug can be seen in both monomers.
(c) Detail of PBP1b interactions with the drug. VR allows for close inspection of specific areas of a structure at any desired angle in an intuitive and straightforward way.
(d) Close-up into MurC catalytic pocket containing the substrate, the cofactor and the two catalytic manganese ions. Areas of a structure away from surfaces are easily accessible through motion controls in VR.
}
\label{roomsnap}
\label{drugsnap}
\label{structsnap}
\label{multisnap}
\end{figure}

Snapshots of the VR application being used and presented here are shown in figures \ref{roomsnap} (a) to (d), where a user was interacting with the four proteins discussed above in a VR environment. Conveying a VR experience in pictures is a difficult task; readers with access to the {\sc HTC Vice} or the {\sc Oculus Rift} can download our preconfigured VR rooms from \cite{MartinsRatamero2017}.
Fig.\ \ref{roomsnap}(a) shows the template room containing the four protein structures . Some physical objects such as cubes and balls have been added to emphasize the sense of ``reality" and physicality around the protein models. This figure shows how is also possible to see both controllers being held by an user, with tool tips added for introducing new players to the controls.
%
%
Fig.\ \ref{structsnap}(b) shows a snapshot of a protein structure being held by an user. In this case, the user is holding up dimeric MraY, with the same structure and coloring as in Fig.\ \ref{mray_struct}(a). A close look to the drug and its interactions with protein residues can be taken as desired by ``grabbing'' the protein with the controller and moving it towards the viewer; tilting the controller allows to rotate the protein structure in an easy and intuitive way.
Further detail on specific residues and drug-binding pockets can be seen by simply approaching the areas of interest, or through clipping the protein structure through the view point. Fig.\ \ref{drugsnap}(c) shows a detail of ampicillin bound to the TP domain of PBP1b. This is the same region presented in light blue in Fig.\ \ref{pbp1b_struct}(d). This figure shows how much and how simple VR allows for in-depth analysis of specific areas of a protein structures.
Finally, in Fig.\ \ref{multisnap}(d), we present a detail of an area of contact between MurC, the substrate and the cofactor, including the two catalytic manganese ions. In the VR environment, investigation of protein-ligand interactions is straightforward, since complete control of the 3D position and rotation of the protein model is mapped to movement of the controllers.

\section{\label{conclusions}Conclusions}
In this work, we present a way to visualise and interact with both static structure and dynamics of proteins by using VR. A software pipeline was constructed that enables non-expert researchers to easily embed protein structures into VR programs using a combination of widely available software and custom-built codes. The immersion and interactivity that VR brings can significantly change the level of details accessible to a researcher when analysing protein-ligand interactions or conformational changes. While it is difficult to convey this with static figures as in Fig.\ \ref{roomsnap}, it is interesting to report our personal experiences with immersive 3D VR. For example, despite having worked with these protein structures for a while and having observed the drug-bound catalytic pockets of these proteins for many times on 2D rendering softwares such as PyMol, only when we observed the same in VR we realized that the 3D configuration of the catalytic pocket was significantly different to the picture conveyed by the 2D projections and that we had built in our minds. These differences could translate into designing more effective modifications into the drug.  

With the decreasing costs for customer VR hardware and an established workflow for importing structures into VR, immersive 3D visualization should become viable for an increasing number of research groups. It is a first step for developing VR-based ways of interacting with proteins and probing their properties. At the moment, there is still the limitation in both 2D and 3D protein visualization software packages that pre-computed conformers are necessary and the interaction between user and structure is only at a non-interactive visualisation level. In the future, physical interactions may become possible, where the user could bend the protein structure or add new functional groups to the ligand while background simulations calculate in real-time whether that changes are mechanically stable or energetically favorable. In such a scenario, immersive VR environments can aid much better than other 2D or 3D visualizations in guiding molecule manipulation.  

As well as VR being a very useful tool for visualisation of both protein conformational changes and drug design, there is also the element of fantastic potentials for outreach and engagement with the general public.

\section{\label{acknowledge}Acknowledgments}
We gratefully acknowledge funding via the EPSRC's \emph{Cross-scale prediction of Antimicrobial Resistance: from molecules to populations} network (EP/M027503/1). UK research data statement: data is available at \cite{MartinsRatamero2017}.


\bibliography{Mendeley_Proteins}

\ifNOSUP\end{document}\else%

\clearpage\newpage
\setcounter{figure}{0}
\setcounter{table}{0}
\def\thefigure{S\arabic{figure}}
\def\thetable{S\arabic{table}}
\setcounter{page}{1}
\pagestyle{plain}

\fi\end{document}